\documentclass[
    ,final            
  ]
  {aipproc}

\layoutstyle{8x11single}

\def\ltsima{$\; \buildrel < \over \sim \;$}
\def\simlt{\lower.5ex\hbox{\ltsima}} 
\def\gtsima{$\; \buildrel > \over \sim \;$}
\def\simgt{\lower.5ex\hbox{\gtsima}} 

\def\deg{\hbox{$^\circ$}}
\def\Chandra{{\it Chandra}}

\begin{document}

\title{Compton X-ray and $\gamma$-ray Emission from Extended Radio Galaxies}


\author{C.~C. Cheung\footnote{Jansky Postdoctoral Fellow. The National
Radio Astronomy Observatory is operated by Associated Universities,
Inc.  under a cooperative agreement with the U.S. National Science
Foundation.}}{address={Kavli Institute for Particle Astrophysics and
Cosmology, Stanford University, Stanford, CA 94305, USA}
}

\begin{abstract} The extended lobes of radio galaxies are examined as sources
of X-ray and $\gamma$-ray emission via inverse Compton scattering of 3K
background photons.  The Compton spectra of two exemplary examples, Fornax A
and Centaurus A, are estimated using available radio measurements in the
$\sim$10's MHz -- 10's GHz range. For average lobe magnetic fields of
$\simgt$0.3--1 $\mu$G, the lobe spectra are predicted to extend into the soft
$\gamma$-rays making them likely detectable with the GLAST LAT.  If detected,
their large angular extents ($\sim$1\deg\ and 8\deg) will make it possible to
``image'' the radio lobes in $\gamma$-rays. Similarly, this process operates
in more distant radio galaxies and the possibility that such systems will be
detected as unresolved $\gamma$-ray sources with GLAST is briefly considered.
\end{abstract}

\maketitle

\section{Inverse Compton "Images" of Large Radio Galaxies}

Inverse Compton (IC) scattering of the CMB is a mandatory process in
synchrotron emitting sources.  This emission becomes most prominent in regions
of weaker $B$-field like the extended lobes of radio galaxies. Many such
IC/CMB lobe X-ray sources are now known (e.g., Croston et al. 2005; Kataoka \&
Stawarz 2005) and we explore the possibility of the IC spectra extending into
the $\gamma$-ray band. This is independent of possible $\gamma$-ray emission
from the unresolved nuclei of radio galaxies, i.e., from the misaligned blazar
(Sreekumar et al. 1999; Bai \& Lee 2001;  Foshini et al. 2005). 

The case of the nearby (D=18.6 Mpc) double-lobed radio galaxy, Fornax A was
discussed in Cheung (2007).  Radio flux density measurements down to $\sim$30
MHz (Isobe et al. 2006) were used to estimate the IC/CMB spectra of the lobes.
Normalizing the IC spectra to the X-ray detections of the lobes (which
indicate $B$$\sim$1.5$\mu$G on average; Feigelson et al.  1995, Isobe et al. 
2006), the presence of high frequency radio emission observed in the
$\simgt$10--90 GHz range with WMAP (with $F_{\nu}\propto\nu^{-1.5}$) imply a
detectable soft $\gamma$-ray signal. As this emission is not expected to be
time variable, the LAT can simply integrate on this position during its normal
scanning mode to test this prediction. 

Here, we similarly consider the case of Centaurus A which is only 3.5 Mpc
away. It is long known to have structure extended over $\sim$8\deg\ in
declination (Cooper et al. 1965, and references therein). We use the extensive
compilation by Alvarez et al. (2000) of the various components of the radio
source; Figure~1 shows a low resolution 408 MHz image from Haslam et al.
(1982).  The outer (degree-scale) giant lobes (GLs) visible in Figure~1
account for $\simgt$2/3 of the total 408 MHz emission at $\sim$1000 Jy each;
the arcmin-scale inner lobes (ILs) are only 3--4 times fainter than each GL. 
The northern GL was searched for such IC emission with ASCA data but the
extended X-rays could not be uniquely attributed to such a process (Isobe et
al.  2001). 

Repeating the analysis as for Fornax~A, it appears that the extended
components of Cen~A will also emit $\gamma$-rays at a level detectable by
GLAST.  The various data from $\sim$10 MHz to 43 GHz are consistent with a
single spectral index $\alpha$=0.7. Since the luminosities of both the ILs
and GLs are similar (within $\sim$20$\%$), only the SEDs of the southern ones
are plotted in Figure~1. Utilizing these radio measurements, the expected
IC/CMB spectra for example $B$-field strengths are drawn.  The integrated
Compton Gamma-Ray Observatory (CGRO) COMPTEL detections of Cen~A (Steinle et
al. 1998) at $\sim$10$^{21}$ Hz already limit $B\simgt 1 \mu$G for both the
northern and southern GLs (since they have similar radio spectra); a similar
extrapolation for the ILs give $B\simgt 0.3 \mu$G. 

Thermal emission will be a complicating factor at energies below $\sim$10 keV,
so hard X-ray and soft $\gamma$-ray measurements are better suited for
detecting the suspected IC/CMB emission.  Additionally, since Fornax A and
Cen~A are quite extended in the sky ($\sim$1\deg\ and 8\deg), if they are
detected with GLAST, the contributions from the two lobes will be separable
with the LAT making IC/CMB $\gamma$-ray ``images'' of these radio galaxies
possible. These $\gamma$-ray images will appear most similar to radio maps
at frequencies, $\nu\simgt$10 GHz;  such radio maps of Cen~A's extended
components have already been obtained by WMAP (Page et al. 2007, Fig.~2
therein) and are available for this comparison.

\section{The Highest-Redshift Radio Galaxies}

Using the above examples as a guide, we can gauge the feasibility of detecting
even more distant radio galaxies at the higher-energies.  Utilizing the recent
large compilation of bright $z$$>$2.5 radio sources by Carson et al.  (2007),
we consider the highest-redshift ($z>3.5$) radio galaxies for illustration.
The observed monochromatic Compton (X-ray, $\gamma$-ray) to synchrotron
(radio) flux ratio for IC/CMB emission has a strong redshift dependence: for
$\alpha$=1, it is simply $f_{\rm c}/f_{\rm s} \simeq u_{\rm cmb}/u_{\rm B}
\simeq 10 (1+z)^{4}\delta^{2}/B^2_{\rm \mu G}$ ($f_{\nu}\equiv\nu$$F_\nu$, and
$\delta$ is the Doppler factor which is set to 1). We use the NVSS (Condon et
al. 1998) 1.4 GHz fluxes for $f_{\rm s}$.  Most of the considered sources
(13/17) are detected at 74 MHz in the VLSS database (Cohen et al. 2005) giving
$\alpha_{\rm 74 MHz - 1.4 GHz}\sim 0.9-1.2$, so the approximate relation is
applicable. 

As in the nearby sources, these distant radio galaxies are expected to be
IC/CMB X-ray sources unless $B\gg$10$\mu$G (Fig.~1).  \Chandra\ observations
should easily detect this emission to constrain the lobe $B$-fields, and thus
the lobe energetics.  In one of the highest-redshift ($z=3.8$) radio galaxies
observed so far with \Chandra\ (Scharf et al.  2003), it was necessary to
remove the contribution from a bright nucleus (spatially) and extended IC
emission from other sources of seed photons (by spectral fitting). Such X-ray
observations can guide our determination of the expected level of (soft)
$\gamma$-ray emission from the IC/CMB process; at the moment, the estimates
(Fig.~1) are rather crude.

\begin{figure*}[t]
  \includegraphics[height=.27\textheight]{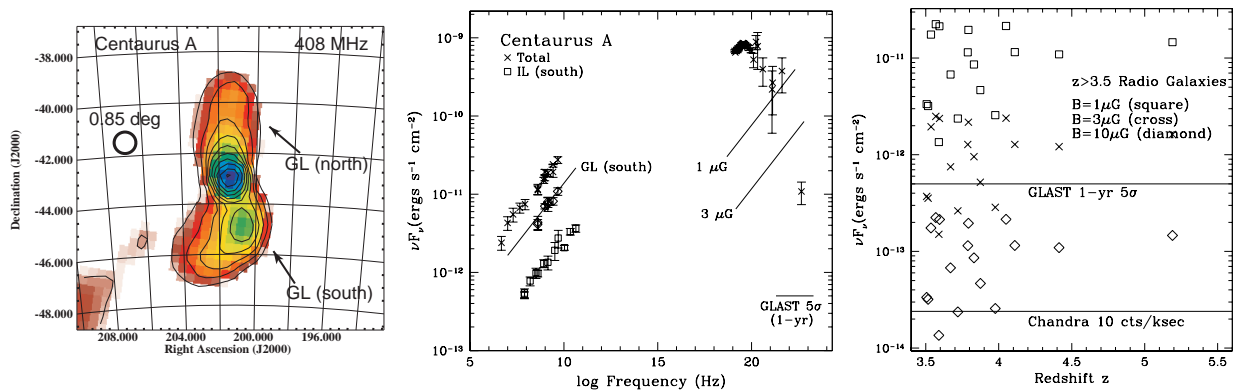}
\caption{[{\it\bf Left}] Radio image of Cen A at 0.85\deg\ resolution which is
comparable to the angular resolution of GLAST/LAT.  [{\it\bf Center}] SEDs of
the multiple components of the Cen~A radio source with lines indicating
$F_{\nu}\propto\nu^{-0.7}$ spectra.  The data points at $>10^{18}$ Hz are the
integrated detections with CGRO with lines indicating the expected IC/CMB
spectra of the southern giant lobe for different average $B$-fields.  [{\it\bf
Right}] IC/CMB X-ray and $\gamma$-ray flux predictions for 17 of the
highest-$z$ radio galaxies discussed in the text. Typical \Chandra\
``snapshots" are 5--10 ksec exposures so these sources are all expected to be
easily detectable in the X-rays for the indicated field strengths. GLAST
detections require electrons with $\gamma\simgt$10$^5$ in a 1$\mu$G or smaller
field which are optimistic. } \end{figure*}

\bibliographystyle{aipproc}   

\bibliography{sample}

\end{document}